# Observation of Highly Nonlinear Resistive Switching of $Al_2O_3$/$TiO_{2-x}$ Memristors at Cryogenic Temperature (1.5 K)

Yann Beilliard[1,2]*, François Paquette[1,2], Frédéric Brousseau[1,2], Serge Ecoffey[1,2], Fabien Alibart[1,2,3], Dominique Drouin[1,2]

[1]Institut Interdisciplinaire d'Innovation Technologique (3IT), Université de Sherbrooke, Sherbrooke, Québec, Canada
[2]Laboratoire Nanotechnologies Nanosystèmes (LN2) – CNRS UMI-3463
[3]Institute of Electronics, Microelectronics and Nanotechnology (IEMN), Université de Lille, Villeneuve d'Ascq, France
*Correspondance: yann.beilliard@usherbrooke.ca – Orcid: https://orcid.org/0000-0003-0311-8840

*Abstract*—In this work, we investigate the behavior of $Al_2O_3$/$TiO_{2-x}$ cross-point memristors in cryogenic environment. We report successful resistive switching of memristor devices from 300 K down to 1.5 K. The I-V curves exhibit negative differential resistance effects between 130 and 1.5 K, attributed to a metal-insulator transition (MIT) of the $Ti_4O_7$ conductive filament. The resulting highly nonlinear behavior is associated to a maximum $I_{ON}/I_{OFF}$ diode ratio of 84 at 1.5 K, paving the way to selector-free cryogenic passive crossbars. Finally, temperature-dependant thermal activation energies related to the conductance at low bias (20 mV) are extracted for memristors in low resistance state, suggesting hopping-type conduction mechanisms.

## I. Introduction

Continuous progress in solid-state quantum technologies has led to promising high-quality silicon-based quantum bits (qubits) [1], [2]. Such quantum systems working at cryogenic temperature down to 10 mK are currently controlled by classical electronics located outside the cryostat at room-temperature. While this approach allows to operate few-qubit systems, it becomes clear that a drastically higher number of qubits would be impossible to manage. Thus, to make the step to large-scale quantum systems, it is necessary to explore novel integration and packaging approaches to develop quantum-classical interfaces in cryogenic environment with one or multiple temperature stages [3]. In the meantime, nanoscale resistive switching memories, also called memristors, are one of the most promising candidates for room-temperature applications such as high-capacity memories and in-memory computing applications based on massively parallel neuromorphic electronic architectures [4]. Demonstrating reversible, non-volatile and highly non-linear resistance programing of memristor devices at cryogenic temperature would paves the way to memristor-based cryogenic electronics, which could help overcome the roadblocks towards quantum supremacy. Up to now, the lowest temperature at which resistive memories were studied is 4 K [5]–[10], mostly to obtain a better understanding of temperature-dependant behavior and conduction mechanisms of devices based on transition metal oxides.

In the present work, we investigate the current-voltage-temperature (I-V-T) characteristics of $Al_2O_3$/$TiO_{2-x}$ cross-point memristors at temperature as low as 1.5 K. Resistive switching cycles in the temperature range of 300-1.5 K are first demonstrated. Then, we discuss the influence of temperature on SET/RESET voltages, I-V characteristics and resistance states. The observation of a negative differential resistance at low temperature from 130 K is also investigated. Finally, the influence of temperature on the low bias conductance and thermal activation energies are examined, along with associated conduction mechanisms.

## II. Experiments

The studied memristors with a TiN/$Al_2O_3$/$TiO_{2-x}$/Ti/Pt structure and a 2×2 µm² area were fabricated through contact UV lithography. The process flow starts with the fabrication of 2 µm wide TiN bottom electrodes (BE) using a back-end-of-line compatible damascene process in $SiO_2$. The switching junction is then fabricated by depositing a 1.4 nm thick layer of $Al_2O_3$ and a 30 nm thick layer of non-stoichiometric $TiO_{2-x}$ by atomic layer deposition (ALD) and physical vapor deposition (PVD) respectively. Finally, 2 µm wide Ti/Pt top electrodes (TE) are patterned using liftoff techniques. Electrical characterizations at ambient and cryogenic temperatures were carried out in a variable temperature insert (VTI) cryostat. A total of 10 devices and test structures were wire bonded to a chip carrier and characterized from 300 to 1.5 K with an Agilent E5270B parametric measurement system. The average resistances of BE and TE were measured at each temperature, allowing the calculation of the voltage $V_{junction}$ seen by the switching junction by subtracting to the total voltage $V_{total}$ the drop of voltage resulting from access resistances.

## III. Results and Discussion

Fig. 1 shows current-voltage characteristics of one of the studied memristors at 300 and 1.5 K. Electroforming was initially performed at ambient temperature with a positive bias applied to the top electrode and a current compliance of 0.6 mA. Successful resistive switching is visible at both room and cryogenic temperatures, with typical progressive SET-RESET behavior of TiO2-based memristors. However, significant differences are noticeable in I-V curves between 300 and 1.5 K. Firstly, the SET and RESET voltages determined by the maximum value of dI/dV appear to vary differently as a function of temperature. As visible in the bottom-right inset of Fig. 1, the SET and RESET voltages are first similar at 300 K, with values of about $V_{SET}$ = 1.08 V and $|V_{RESET}|$ = 1.05 V. However, only the RESET voltages tend to be significantly affected by temperature with a maximum value of $|V_{RESET}|$ = 1.28 V at 1.5 K, which corresponds to an increase of 0.25 V compared to 300 K. Regarding the SET voltages, the values are mostly stable around $V_{SET}$ = 1.1 V as temperature decreases. The asymmetrical variations of $V_{SET}$ and $|V_{RESET}|$ with temperature could be related to the different temperature dependencies of the SET and RESET processes, involving oxygen vacancies migration and redox mechanisms [11]. On the one hand, the growth of the conductive filament during the SET step is mainly due to electrochemical reduction of the conductive filament followed by the migration of consequent oxygen vacancies induced by the electric field [12] and the Soret effect associated to local temperature gradients [13]. On the other hand, several studies describe the RESET process as temperature-activated migration of oxygen ions and oxygen vacancies, leading to the rupture of the conductive filament and thus a decrease in conductance [11], [14], [15]. Therefore, a higher RESET bias must be applied as temperature decreases to generate enough Joule heating to trigger ionic migration.

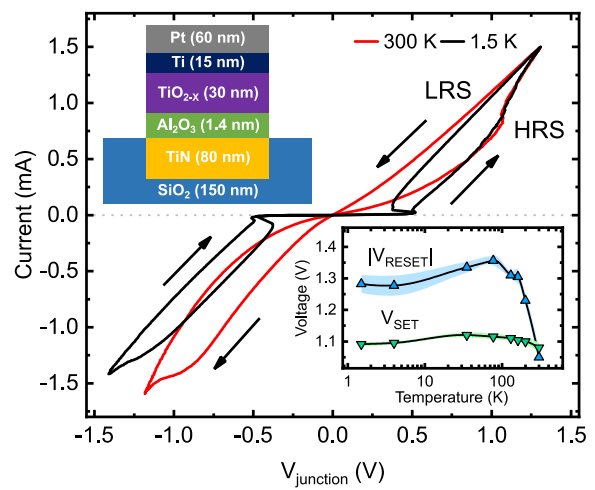

Fig. 1. Experimental resistive switching at 300 K (red line) and 1.5 K (black line). Negative differential resistance is visible at around 0.5 V. The top-left inset is a schematic cross-section of the cross-point memristors studied. The bottom-right inset shows the evolution of $V_{SET}$ and $|V_{RESET}|$ values as a function of temperature in semi-log scale.

The second major observable difference between 300 and 1.5 K is the high nonlinearity of the I-V curve in cryogenic conditions, exhibiting a negative differential resistance (NDR) behavior around 0.5 V. To further investigate this temperature-induced phenomenon and its interplay with non-volatile resistive switching, Fig. 2 shows I-V characteristics of a memristor in low resistance state (LRS) at various temperatures, with the total voltage $V_{total}$ in abscissa. Note that between each temperature stage, a full SET/RESET cycle was performed. We can observe a decrease in conductance with decreasing temperature, which is expected for metal-oxide-metal structures. Starting from 130 K and down to 1.5 K, highly nonlinear I-V characteristics are clearly noticeable, with NDR effects during current-controlled positive sweeps and sudden jumps in current during voltage-controlled negative sweeps. This threshold switching behavior is volatile and seems to be only related to temperature- and voltage-driven electronic transport properties of the conductive filament. It has been demonstrated that the conductive filament of TiO2-based memristors could be composed of sub-oxides phases such as Ti4O7 [16], [17]. This Magnéli phase of titanium oxide is known to undergo an abrupt metal-insulator transitions (MIT) at temperatures ranging from 155 down to 120 K [18], [19]. We therefore attribute the drastic modification of the I-V curves around 130 K to MIT of the Ti4O7 domains inside the conductive filament induced by Joule heating. The threshold voltages $V_{th}$ visible in Fig. 2 thus correspond to the transition between metallic and insulating regimes. The values of $V_{th}$ being directly related to the geometry and composition of the conductive filament, the fact that $V_{th}$ values increase with decreasing temperature is due to the higher power needed to heat up the Ti4O7 to the critical temperature range of 120-155 K corresponding to MIT [19].

While the addition of oxides such as VO2 or NbO2 to the switching stack is currently investigated to obtain room temperature MIT-based access devices in memristor arrays [20], the high nonlinearity of our Al2O3/TiO2-x memristors at cryogenic temperature induced by the thermoelectrically triggered MIT of Ti4O7 paves the way to selector-free cryogenic passive crossbar architectures. In that scope, we calculated the $I_{ON}/I_{OFF}$ diode ratio as a function of temperature, measured from Fig. 2 with $I_{ON}$ and $I_{OFF}$ currents evaluated at $V_{READ}$ = 0.6 V and $V_{READ}$/2 respectively. The obtained values range from 2 to 84 between 300 and 1.5 K, with a significant increase after 130 K. Furthermore, in order to evaluate the influence of temperature and thus MIT on the resistive switching process, the average resistances in low resistance state $R_{LRS}$ and high resistance state $R_{HRS}$ were measured at $V_{READ}$ = 0.6 V over multiple switching cycles. Fig. 3 shows the results along with $R_{HRS}/R_{LRS}$ values as a function of temperature. Two regimes can be distinguished, defined by the critical temperature of 130 K associated with the onset of MIT for our devices. When the filament is in metallic regime, $R_{HRS}/R_{LRS}$ has a relatively constant value of around 1.35, despite the slight increase of both $R_{HRS}$ and $R_{LRS}$ related to the temperature dependence of the conduction mechanism. The insulating regime of the filament exhibits a slight increase of $R_{HRS}/R_{LRS}$ up to 1.65 and higher variability of $R_{HRS}$ values at the lowest temperatures. As previously discussed, the rupture of the conductive filament during the RESET step is mainly a temperature-driven mechanism. In cryogenic conditions, only the Joule heating contributes to this process, which causes inherent fluctuations in the filament geometry. This eventually results in variations of the conductance values in HRS from one cycle to the other.

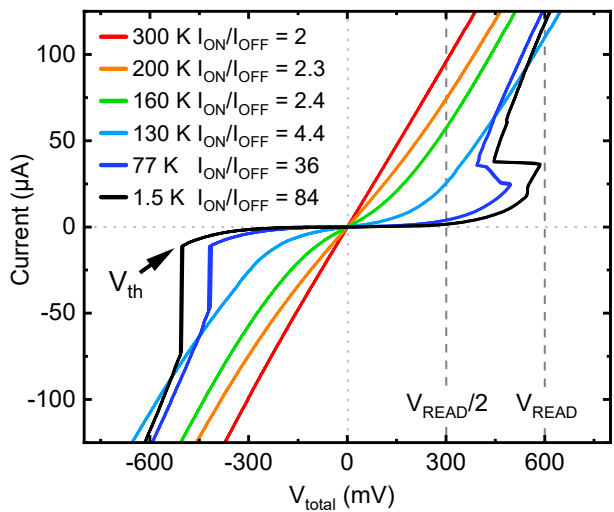

Fig. 2. I-V curves of a memristor in LRS between 300 and 1.5 K, with current-controlled positive sweeps and voltage-controlled negative sweeps. Negative differential resistance effects attributed to a metal-insulator transition appear around 130 K. $I_{ON}/I_{OFF}$ diode ratios measured at $V_{READ}$ = 0.6 V and $V_{READ}$/2 are indicated. At 1.5 K, the $I_{ON}/I_{OFF}$ ratio is increased by a factor 42 compared to 300 K.

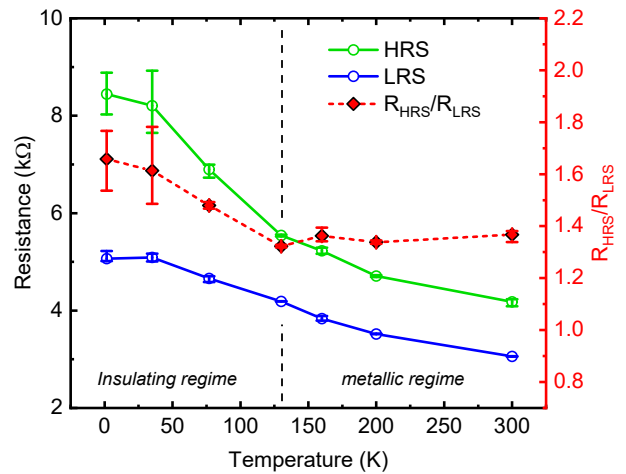

Fig. 3. Average resistance values of a device in LRS and HRS measured at $V_{READ}$ = 0.6 V as a function of temperature. $R_{HRS}/R_{LRS}$ ratio values are plotted in red. Error bars represent minimum and maximum resistance and ratio values over multiple switching cycles at a given temperature.

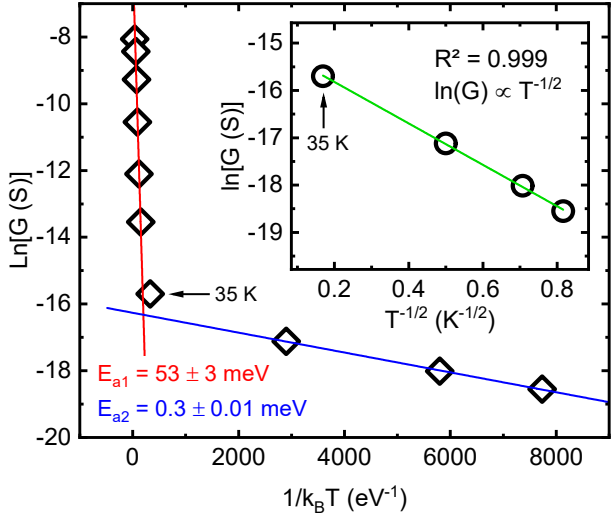

Fig. 4. Arrhenius plot of the low bias conductance measured at V = 20 mV between 300 and 1.5 K for a memristor in LRS. The activation energies $E_{a1}$ and $E_{a2}$ are directly extracted from the slope of the linear fits. The inset shows the linear fitting of $\ln(G)$ versus $T^{-1/2}$ in the temperature range of 35-1.5 K, suggesting an Efros-Shklovskii variable-range hopping conduction mechanism.

This next section is dedicated to gain a better understanding of the influence of temperature on the conduction mechanism at low bias, i.e. below non-volatile resistive switching regime and MIT. As observed in Fig. 2, the conductance $G$ decreases with decreasing temperature $T$. In general, the temperature-dependent conductance $G(T)$ of disordered insulating thin films such as $Al_2O_3/TiO_{2-x}$ can be described by

$$G(T) \propto \exp(-T_0/T)^\alpha \quad (1)$$

where the temperature exponent $\alpha$ depends on the transport mechanism and $T_0$ is the characteristic temperature [21], [22]. In order to determine the value of $\alpha$, Fig. 4 shows the Arrhenius plot of the low bias conductance of a device in LRS, measured from I-V-T curves at 20 mV. A change of slope is clearly noticeable as temperature decreases, defining two different regimes that we can associate to the metallic and insulating states of the conductive filament. Between 300 and 77 K, the data fit linearly with the Arrhenius equation

$$G(T) \propto \exp(-E_a/k_BT) \quad (2)$$

where $E_a$ is the thermal activation energy and $k_B$ is the Boltzmann constant. This implies that $\alpha = 1$ in that temperature range, which can correspond to a nearest neighbor hopping (NNH) conduction mechanism between localized states [22]. The extracted thermal activation energy of the hopping electrons is $E_{a1} = 53 \pm 3$ meV, which is in agreement with recent work on $TiO_x$-based memristors [9]. Regarding the insulating regime, while no activation energy can usually be found for pure $Ti_4O_7$ [23], we extracted an activation energy of $E_{a2} = 0.3 \pm 0.01$ meV. Such low energy value implies that the NNH behavior is replaced by a weakly temperature-dependant conduction mechanism at cryogenic temperature. Based on equation (1), we found that $\ln(G)$ varies linearly with $T^{-1/2}$ between 35 and 1.5 K, as depicted in the inset of Fig. 4. The fact that $\alpha = 1/2$ suggests that below 35 K the transport mechanism is dominated by Efros-Shklovskii variable-range hopping (ES-VRH) [24]. In this case, the hopping occurs under the influence of long-range electron-electron interactions between localized electrons, creating a Coulomb gap in the density of states near the Fermi energy [21], [24].

## IV. CONCLUSION

Electrical characterizations of $Al_2O_3/TiO_{2-x}$ memristors in cryogenic conductions have been performed, demonstrating successful resistive switching at temperature as low as 1.5 K. We suggest that the asymmetrical variation of SET and RESET voltages as temperature decreases is due to the high temperature dependence of the RESET process. The negative differential resistance effects observed in the temperature range of 130-1.5 K is attributed to a metal-insulator transition of the $Ti_4O_7$ conductive filament. The resulting highly nonlinear behavior of the current-voltage characteristics exhibit a $I_{ON}/I_{OFF}$ diode ratio of 84 at 1.5 K, offering opportunities for selector-free memristor crossbar development for cryogenic electronics. However, variations in resistance due to fluctuations of the filament geometry between switching cycles seem to be accentuated at low temperature, which should be considered when designing and operating such crossbar-based system. Finally, at low bias and in low resistance state, the study of the temperature dependence of the conductance suggests that hopping is the dominant transport mechanism, with nearest neighbor hopping from 300 to 77 K and Efros-Shklovskii variable-range hopping between 35 and 1.5 K.

ACKNOWLEDGMENT

This research was funded by Natural Sciences and Engineering Research Council of Canada (NSERC). We acknowledge financial supports from the EU: ERC-2017-COG project IONOS (# GA 773228).